\documentclass[sigconf]{acmart}
\usepackage{booktabs} 
\usepackage[normalem]{ulem}
\usepackage{multirow}
\usepackage{bm}
\useunder{\uline}{\ul}{}
\usepackage{subcaption}
\usepackage{enumitem}
\usepackage[para]{footmisc}
\usepackage[T1]{fontenc}
\hypersetup{
,bookmarksnumbered=true%
,hypertexnames=false%
,breaklinks=true%
,colorlinks=true%
,linkcolor=black%
,citecolor=black%
,urlcolor=black%
}

\definecolor{tealblue}{rgb}{0.21, 0.46, 0.53}
\definecolor{wildstrawberry}{rgb}{1.0, 0.26, 0.64}
\definecolor{ao(english)}{rgb}{0.0, 0.5, 0.0}

\copyrightyear{2019}
\acmYear{2019}
\setcopyright{acmcopyright}
\acmConference[CHIIR '19]{2019 Conference on Human Information Interaction and Retrieval}{March 10--14, 2019}{Glasgow, Scotland Uk}
\acmBooktitle{2019 Conference on Human Information Interaction and Retrieval (CHIIR '19), March 10--14, 2019, Glasgow, Scotland Uk}
\acmPrice{15.00}
\acmDOI{10.1145/3295750.3298946}
\acmISBN{978-1-4503-6025-8/19/03}

\def\ignore#1{}

\begin{document}
\title{Answer Interaction in Non-factoid Question Answering Systems}

\begin{abstract}
Information retrieval systems are evolving from document retrieval to answer retrieval. Web search logs provide large amounts of data about how people interact with ranked lists of documents, but very little is known about interaction with answer texts. In this paper, we use Amazon Mechanical Turk to investigate three answer presentation and interaction approaches in a non-factoid question answering setting. We find that people perceive and react to good and bad answers very differently, and can identify good answers relatively quickly. Our results provide the basis for further investigation of effective answer interaction and feedback methods.

\end{abstract}





\author{Chen~Qu,~Liu~Yang,~W.~Bruce~Croft}
\affiliation{
	\institution{University of Massachusetts Amherst}
}
\email{{chenqu, lyang, croft}@cs.umass.edu}

\author{Falk Scholer}
\affiliation{
	\institution{RMIT University}
}
\email{falk.scholer@rmit.edu.au}

\author{Yongfeng Zhang}
\affiliation{
	\institution{Rutgers University}
}
\email{yongfeng.zhang@rutgers.edu}


\affiliation{%
  \institution{}
  \streetaddress{}
  \city{}
  \state{}
  \postcode{}
}

\maketitle

\section{Introduction}
\label{sec:intro}

Classic information retrieval (IR) systems aim to return a list of relevant documents on a search engine result page (SERP). This type of presentation is often described as ``ten blue links'', because users typically need to click on the ranked results and be redirected to the documents. Modern search engines have paid attention to search results diversification~\cite{Agrawal2009Diverse, Santos2015Diverse} and heterogeneous content presentation~\cite{Wang2016BeyondRanking}. Recently, several works have focused on retrieving extractive answers instead of documents~\cite{Yih2013FactoidQA, Yu2014FactoidQA, Keikha2014NonFactoidQA, Park2015NonFactoidQA,Yang2016NonFactoidQA}. Industrial examples include Google's \textit{featured snippets},\footnote{https://support.google.com/webmasters/answer/6229325} which display a potential answer extracted from the top search result.

If search engines can return a list of potential answers rather than documents, it is essential to study the most effective way to present these answers and interact with users. Specifically, this research question should be emphasized in non-factoid question answering (QA) systems. This is because non-factoid QA poses unique challenges to answer presentation and interaction as it requires several answer sentences or passages, instead of simple entity-based answers as in factoid QA.

User interaction with SERPs has been widely studied using search logs that contain clicks and query reformulations~\cite{Agichtein2006QueryLog, Silverstein1999QueryLog}. Furthermore, other works focus on interaction and feedback methods for document retrieval by studying real users instead of search logs~\cite{Kelly2001Interaction, Koenemann1996Interaction}. However, fine-grained presentation and interaction processes with answers have rarely been investigated in previous work. Additional information is needed from observing what constitutes a good answer when users provide fine-grained and precise feedback, instead of simply indicating whether the answer is relevant or not. We believe that studying fine-grained user interaction and feedback can lead to more effective answer finding, as well as having an impact on the design of conversational search systems. 

In this work, we investigate three answer presentation and interaction approaches (Line by Line, Passage Highlight, and Passage Highlighting with Suggested Words) to understand how people perceive good and bad answers. The Line by Line setting reveals a potential answer passage one line at a time and observes people's reactions as they go through the passage. The Passage Highlight setting presents the full passage, and instructs users to highlight important words that make them believe the passage is a good or bad answer. The third setting is built upon the second one, and includes some suggested words emphasized with special styles. We hired crowdsourcing workers from Amazon Mechanical Turk (MTurk)\footnote{\url{https://www.mturk.com/}} to conduct the experiments. Based on these fine-grained experiments, we find that people perceive good answers and bad answers very differently, which could lead to more effective relevance feedback schemes. For example, people do not hesitate to rate a bad answer, but they can be severe on the answer quality judgments even in some cases where the passage is the answer. Another finding is that people's initial impressions of answer quality are usually correct, and they become more and more confident about answer quality as they go through the answer. In addition, we investigate the relation between answer quality and QA text similarity and find that they are not always correlated.

Our contributions can be summarized as follows. (1) We conduct one of the first fine-grained analyses on answer presentation and interaction in a non-factoid QA setting. (2) We provide an empirical analysis to answer an important research question: is answer quality related to QA text similarity? Our findings can be used to design a more interactive IR system that emphasizes answer retrieval. In addition, our work also has implications for conversational search, since it is essentially a multi-turn interaction process.

\section{Related Work}
\label{sec:relatedwork}
\textbf{User Interaction and Relevance Feedback}. Relevance feedback~\cite{Lv2009RF, Diaz2008RF, Zhai2001RF, Brondwine2016RF, Kelly2001Interaction} is an important and early interactive method in IR systems. In practice, pseudo-relevance feedback (PRF)~\cite{Lv2010PRF, Lavrenko2001PRF, Collins-Thompson2007PRF} is widely used. It assumes that top-ranked documents are relevant and thus can be used for query expansion. In contrast to PRF, some approaches focus on explicit interactions with the user~\cite{Oddy1977Information, Croft1987Feedback, Aalbersberg1992Feedback, Koenemann1996Interaction}.
This work builds on these early papers and focuses on explicit and fine-grained user interaction with answers. Our methods could be especially useful when the interaction bandwidth is limited, such as in mobile search and conversational search.

\textbf{Answer Retrieval}. IR systems are evolving from document retrieval to answer retrieval. Substantial work has been done in factoid QA~\cite{Severyn2015FactoidQA, Yu2014FactoidQA, Yih2013FactoidQA, Yang2016FactoidQA, Iyyer2014FactoidQA}, community QA~\cite{Xue2008CQA, Yang2013CQA, Jansen2014CQA, Surdeanu2008CQA, Surdeanu2008CQA, Surdeanu2011CQA}, and non-factoid QA~\cite{Park2015NonFactoidQA, Cohen2016NonFactoidQA, Keikha2014NonFactoidQA,Yang2016NonFactoidQA}. All these works focus on finding effective methods for answer retrieval. However, even the most effective method can occasionally fail to find the answers. In that case, it is essential to employ user interaction and feedback methods to retrieve the answer in an iterative manner. In this work, we study answer presentation and interaction techniques in a non-factoid QA setting, as an essential complement to answer retrieval models. 

\textbf{Information-seeking Conversations}. An information-seeking conversation typically involves multiple turns of interaction and information exchange between an information seeker and provider. Radlinski and Craswell~\cite{Radlinski2017ConvSearch} described a theoretical framework for conversational search and desirable properties in such systems. In addition, several works~\cite{Trippas2018ConvSearch, thomas2017MISC, Qu2018AnalyzingAC, Wu2003InfoSeeking, Wu2003InfoSeeking2} observed and studied information-seeking conversations between humans and addressed different facets of such interactions. Finally, conversational recommendation~\cite{Zhang2018Rec} and response ranking~\cite{Yang2018RespRank} have been explored under this multi-turn interaction setting.
\section{Our Approach}
\label{sec:approach}
\subsection{Overview}
\label{subsec:overview}
We conduct an observational study of how people perceive answer quality under different interactive settings. This can help us identify effective methods for answer presentation and interaction. In this task, people are given a question and a short passage. The passage may or may not be a good answer to the question.\footnote{``Good'' refers to good quality. Verifying the facts in the passages is not required.} We present the answer passage in three ways, namely, Line by Line, Passage Highlight, and Passage Highlighting with Suggested Words. These settings are designed to obtain fine-grained user feedback to test various answer presentation and interaction methods.

\subsection{Line by Line}
\label{subsec:line-by-line}
In this setting, the answer passage is presented line by line. One line is typically one sentence. At each line, we instruct the annotators to give a rating of how confident they are that the passage is or contains a good answer to the question. This rating is based on the current line and previous lines. Lines that follow the current line are hidden. The confidence rating is provided on a scale of -2 to 2:
\begin{itemize}[leftmargin=*, noitemsep, topsep=0pt]
\item -2: Confident this passage is not an answer to the question.
\item -1: Believe that this passage might not be an answer.
\item 0: Not sure yet.
\item 1: Believe that the passage might be an answer to the question.
\item 2: Confident this passage is an answer to the question.
\end{itemize}
This setting is designed to observe the evaluation of answer quality as people go through a potential answer.

\subsection{Passage Highlight}
\label{subsec:passage-highlight}
In this setting, we present the full answer passages and instruct the annotators to highlight positive and negative words or phrases in the passage (sentences are not encouraged). The \textit{positive words} are those that help to convince the annotators that the passage is a good answer. For example, these words may present a specific answer or introduce key arguments. In contrast, the \textit{negative words} make the annotators feel the answer is of bad quality. For example, these words can be indicators of irrelevant issues, or may reveal that the answer providers are uncertain about their answers. Annotators are instructed to highlight complete words only. At least one highlight for each passage needs to be made for a successful submission. Figure~\ref{fig:interface-example} gives an illustration for the highlighting interface.\footnote{The words in blue are marked for the next setting. They are in black in this setting.} In addition, annotators are asked to give a rating on overall answer quality. The answer quality can be chosen on a scale of 0 to 2:
\begin{itemize}[leftmargin=*, noitemsep, topsep=0pt]
\item 0: It is not an answer to the question.
\item 1: It is an answer to the question, but not of good quality.
\item 2: It is a good answer to the question.
\end{itemize}
This setting is designed to obtain fine-grained feedback on answer quality evaluation and observe rating agreement.

The main challenge of this setting is to quantify annotators' agreement on highlights. Although the annotators are instructed to highlight whole words only, they sometimes highlight partial words. So first, the character-level highlights need to be transformed into word-level highlights. Then a \textit{consistency rule} is applied to highlight all occurrences of a word if this word is highlighted by the annotator. This rule is not applicable to stop words. Then the next step is transforming the highlighted passage into a string by denoting the highlighted words as ``1'' with others as ``0''. In this way the agreement of highlights can be cast into a string similarity problem. The overlap coefficient~\cite{overlap_coef} is used to tackle this problem: 
\begin{equation}\label{eqn:overlap-coef}
\footnotesize
\begin{aligned}
\text{overlap}(H_1, H_2) = \frac{|H_1 \cap H_2|}{\text{min}(|H_1|, |H_2|)}
\end{aligned}
\end{equation}
where $H_1$ and $H_2$ are string representations of highlights from two annotators. This method can compute agreement among multiple annotators. However, since complete agreement with more than two annotators is rare due to the open style of the task, only pairwise agreement is computed. 
The largest value among all pairwise agreements is considered as the agreement for this QA pair. We adopt this setting because perceptions and highlights of answer key words are highly subjective.

\subsection{Passage Highlighting with Suggested Words}
This setting is very similar to the previous setting. The only difference is that we mark some suggested words in the passage for the annotators' reference. These suggested words are marked in bold font and blue color to draw user attention. Figure~\ref{fig:interface-example} gives an example of the highlighting interface. The suggested words meet one of the following criteria: (1) Words that start with a capital letter, such as  acronyms or proper nouns. Words that start a sentence have been excluded. (2) Words that have the top five tf-idf value in this passage. Idf values are computed with a Wikipedia dump (date: 20180520). The annotators understand that they do not have to keep to the suggested words. This setting is designed to compare reactions with and without the suggested words.

\section{Experiments}
\label{sec:experiments}
\subsection{Data Preparation}
\label{subsec:data}

We sampled 200 QA pairs from nfL6,\footnote{https://ciir.cs.umass.edu/downloads/nfL6/} a non-factoid community QA dataset.
A question typically comes with multiple answers provided by the community, with one of them selected as the accepted answer. Our sampled data contains 100 questions, and each question has a good answer and a bad answer. The good answers are the accepted answers of the questions. To match a bad answer to each question, we first use BM25 to collect a small pool of candidate answers for each question and then use a BLSTM model~\cite{Cohen2016NonFactoidQA} to rerank.
We set a criterion that all the answers can be naturally split into four sentences so that the answers would have appropriate lengths, and to work with a fixed number of lines in the Line by Line setting. 

\subsection{MTurk Setup}
\label{subsec:mturk-setup}
We employed crowdsourcing workers (turkers) through Amazon Mechanical Turk (MTurk) to annotate the QA pairs. The three settings are conducted separately. Each QA pair receives annotations from three different turkers. Turkers conduct annotation in the form of assignments, which contain instructions, annotation examples, quiz questions on the instructions, and five QA pairs to annotate (with a combination of good and bad answers). 
We only use annotations from turkers who have passed the quiz test in the analysis. In addition, the turkers are required to have a HIT (Human Intelligence Task) approval rate of 95\% or higher, a minimum of 1,000 approved HITs, and be located in US, Canada,
Australia or Great Britain. The turkers are paid \$0.5/assignment. 

\subsection{Line by Line}
\label{subsec:line-by-line-exp}
\begin{figure}[ht]
	\centering
	\begin{subfigure}[b]{0.45\textwidth}
		\includegraphics[width=\textwidth]{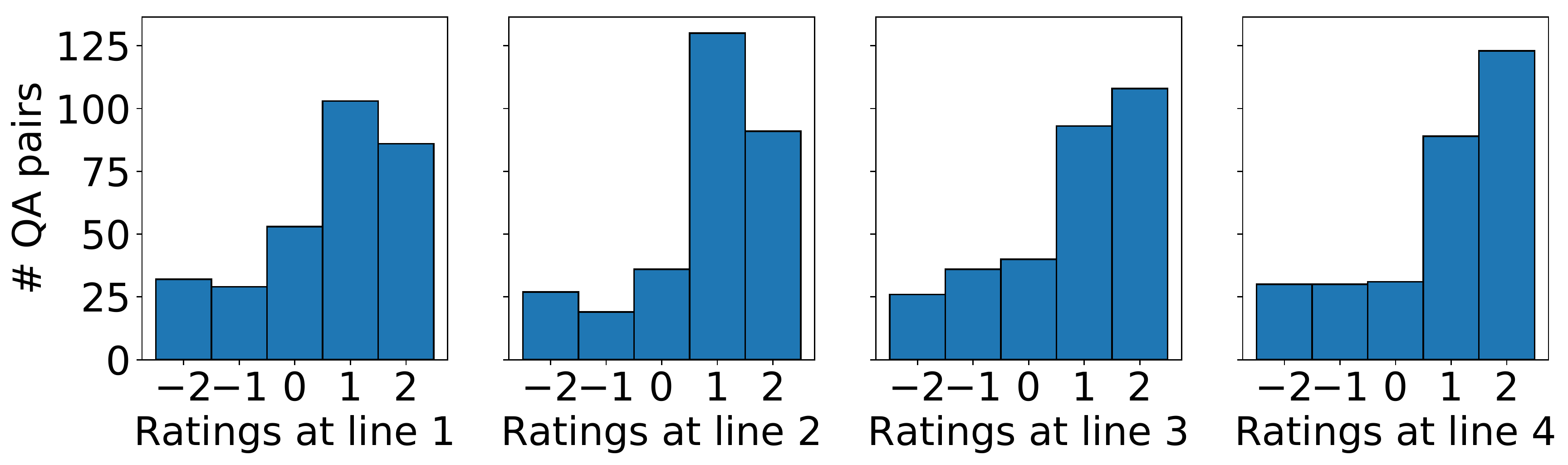}
		\vspace{-0.7cm}
		\caption{Good}
		\label{fig:line-by-line-ratings-good}
	\end{subfigure}
	
	\begin{subfigure}[b]{0.45\textwidth}
        \includegraphics[width=\textwidth]{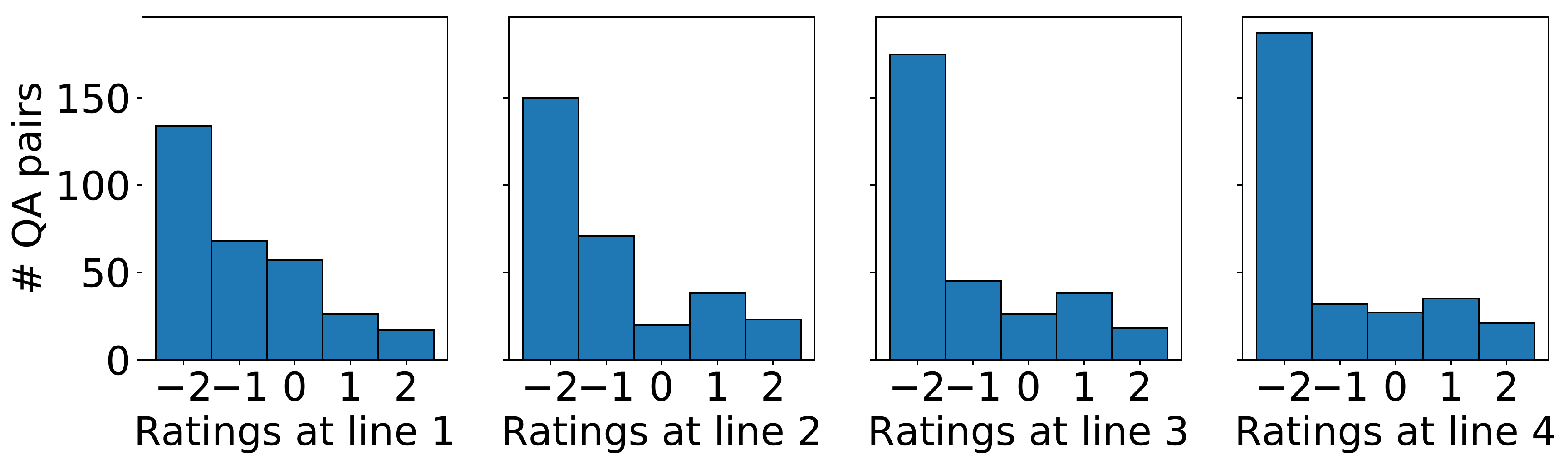}
		\vspace{-0.7cm}
		\caption{Bad}
		\label{fig:line-by-line-ratings-bad}
	\end{subfigure}
	\vspace{-0.5cm}
	\caption{Distribution of confidence ratings from line 1 to 4}
	\label{fig:dist-confidence-ratings}
\end{figure}

Figure~\ref{fig:line-by-line-ratings-good} presents the distribution of confidence ratings from line 1 to line 4 for good answers. The most common confidence rating for line 1 and line 2 is 1.
The most common rating gradually shifts to 2 at line 3 and line 4. 
The distribution of ratings consistently moves up to 2 from line 1 to line 4.
Figure~\ref{fig:line-by-line-ratings-bad} presents the distribution of confidence ratings from line 1 to line 4 for bad answers. The most common confidence rating is always -2 through line 1 to line 4.
The number of -2 ratings gradually increases as the passage is revealed to the turkers.
These indicate that, \textit{for good answers, the turkers have a sense that the answers might be good at the beginning, but they hesitate to make confident ratings until the latter half of the passage is revealed. In contrast, for bad answers, most of the turkers are able to determine the answer quality from the beginning. } 

We also use a $\chi^2$ test to evaluate the difference of confidence ratings between the previous line (expectation) and the current line (observation). We observe the same patterns for good answers and bad answers: the shifts of confidence ratings from line 1 $\rightarrow$ 2 and line 2 $\rightarrow$ 3 are statistically significant with p-value \textless 0.01, while line 3 $\rightarrow$ 4 shows an insignificant difference. This indicates that \textit{some people can determine answer quality quickly while others are slower, but they can make a decision before the last line is revealed}.

\begin{figure}[ht]
\centering
\begin{minipage}{0.238\textwidth}
  \centering
  \includegraphics[width=0.98\linewidth]{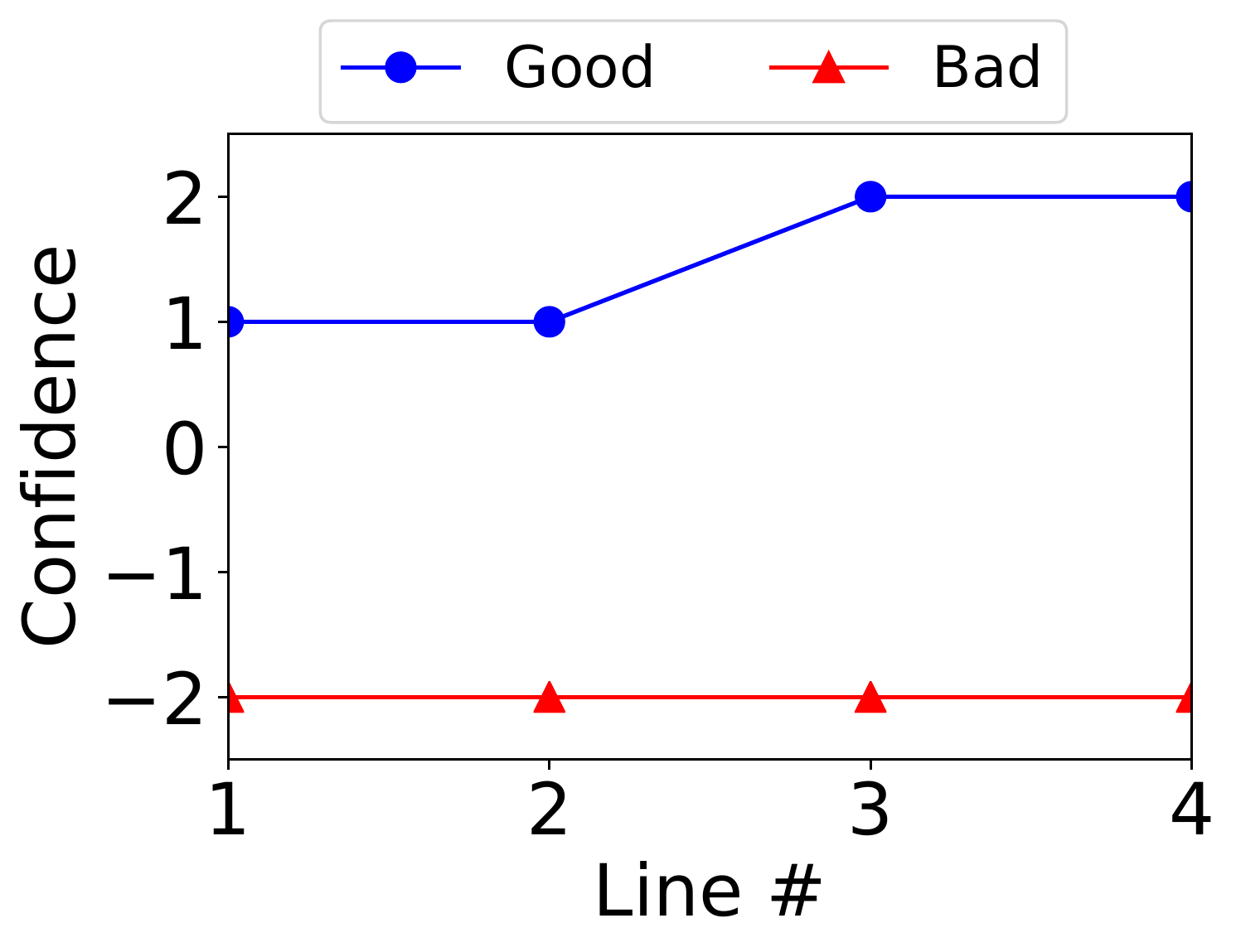}
  \vspace{-0.5cm}
    \caption{Overall confidence}
	\label{fig:line-by-line-overall-majority}
\end{minipage}%
\begin{minipage}{0.238\textwidth}
  \centering
  \includegraphics[width=0.98\linewidth]{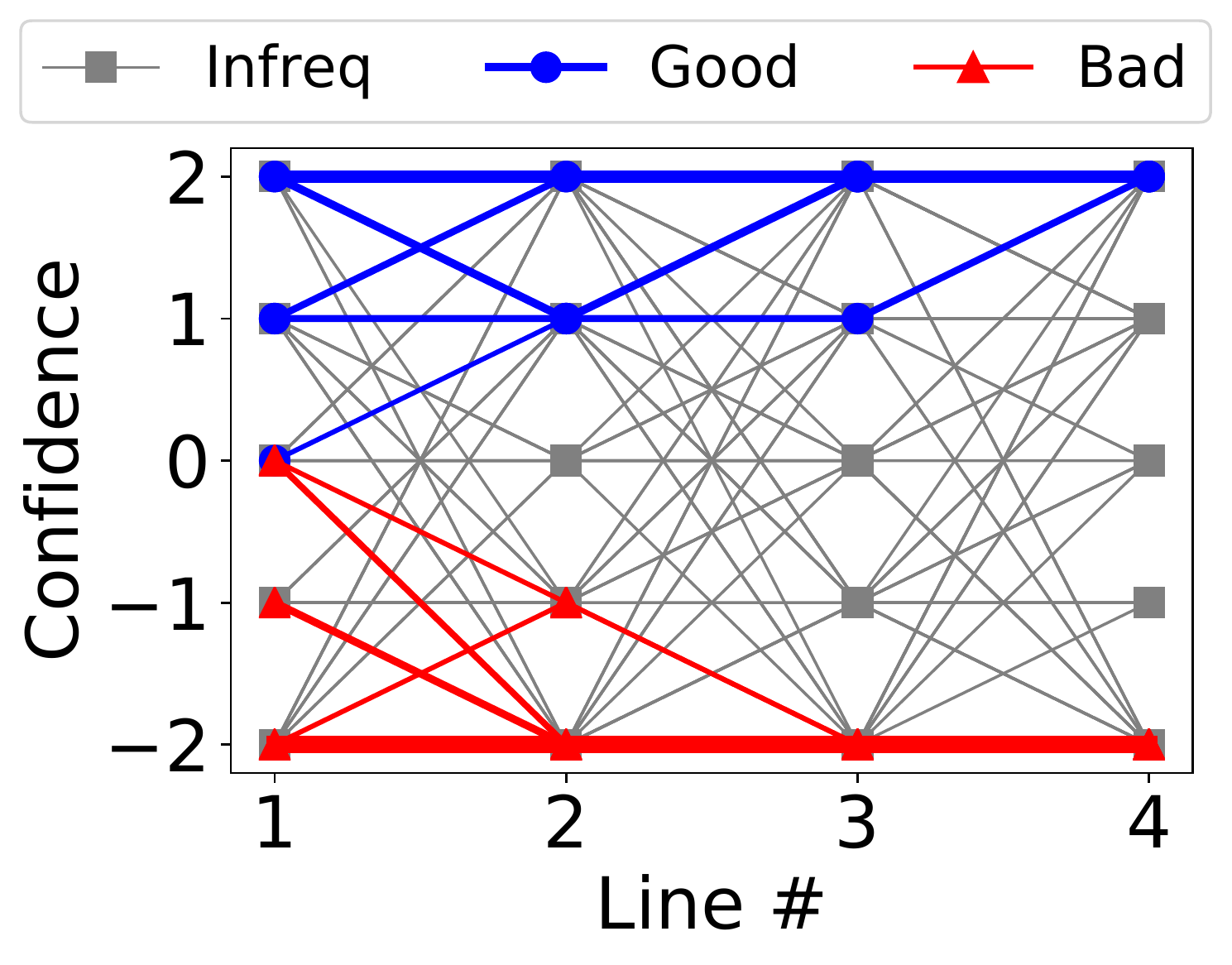}
  \vspace{-0.5cm}
	\caption{Common trends}
	\label{fig:line-by-line-common-trends}
\end{minipage}
\end{figure}

Figure~\ref{fig:line-by-line-overall-majority} presents the majority of confidence ratings for all questions. The rating leaps from 1 to 2 between line 2 and line 3 for good answers. The rating remains at -2 throughout the passage for bad answers. This result double confirms the conclusions above.

In addition to the overall analysis presented above, we also focus on the level of individual QA pairs. Since each QA pair is annotated by three different turkers, we take the majority vote of the ratings at each line as the final rating. We plot the rating trends of individual QA pairs for a better illustration in Figure~\ref{fig:line-by-line-common-trends}. We only plot the most commonly observed trends with more than two occurrences (in blue or red) to reduce noise. For example, if four QA pairs have the same trend of ``$0 \rightarrow 1 \rightarrow 2 \rightarrow 2$'', we consider the occurrence of this trend as four. These common trends constitute about half of total trends. We also plot the remaining trends in gray to show that infrequent trends can be very diverse. The line widths are set to the square root of the trend occurrences to demonstrate trend frequency, while avoiding lines being too wide. 

As presented in the figure above, the common answer quality ratings start at (0, 1, 2) and converge to 2 for good answers. For bad answers, common ratings start at (0, -1, -2) and converge to -2. In addition, a large portion of good answers have consistent ratings of 2 at all lines. Similarly, a very common trend for bad answers is -2 at all lines. These observations indicate that \textit{initial impressions of answer quality are usually correct, because good answers have positive ratings and bad answers have negative ratings from the very beginning. In addition, people become more and more confident about answer quality as they go through the answer}. 

\subsection{Passage Highlight}
\label{subsec:passage-highlight-res}
\begin{figure}[ht]
\centering
\begin{minipage}{0.225\textwidth}
  \centering
  \includegraphics[width=1\linewidth]{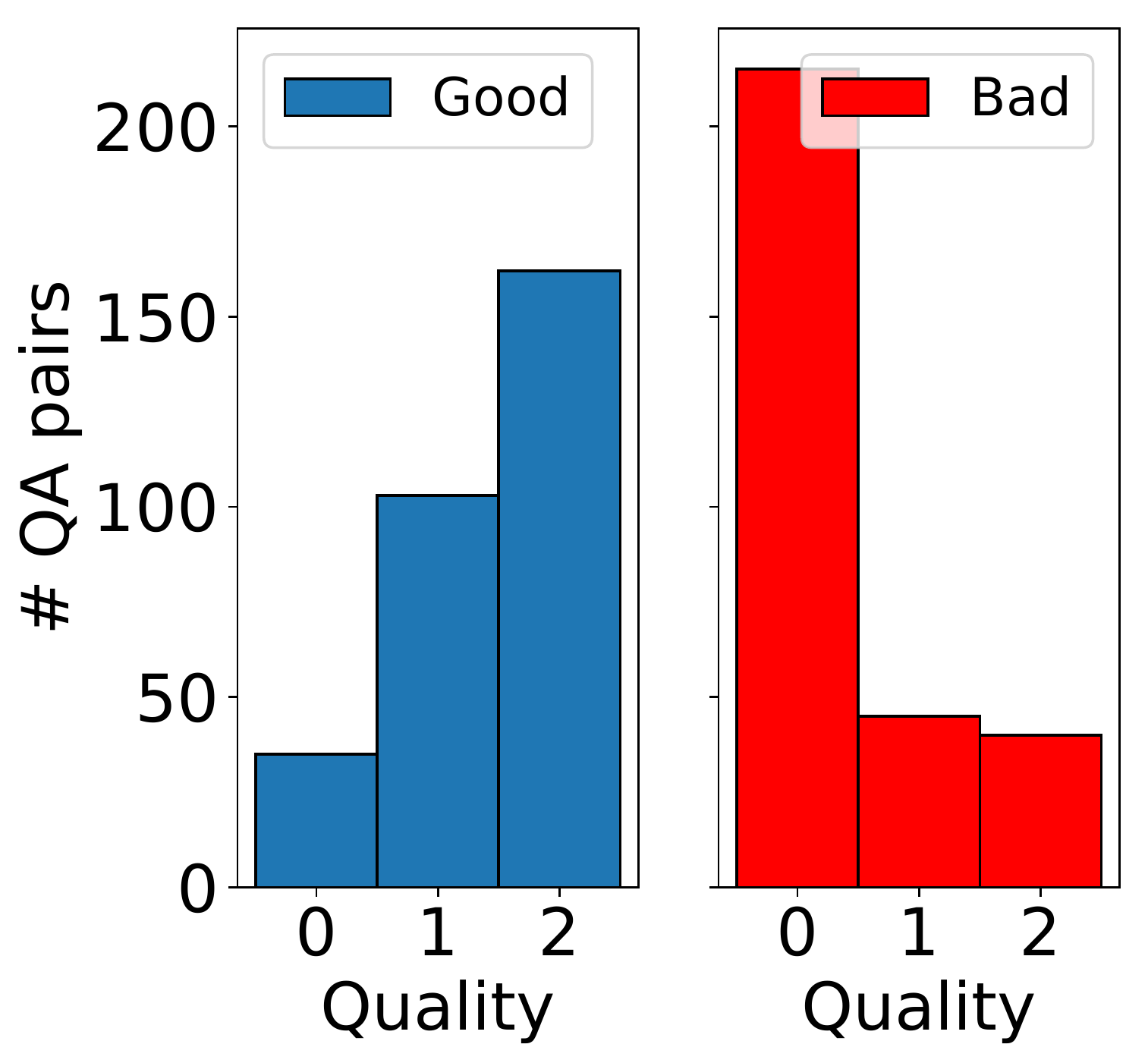}
  \vspace{-0.85cm}
    \caption{Confidence ratings for the second setting}
	\label{fig:passage-highlight-ratings}
\end{minipage}%
\hspace{0.2cm}
\begin{minipage}{0.225\textwidth}
  \centering
  \includegraphics[width=1\linewidth]{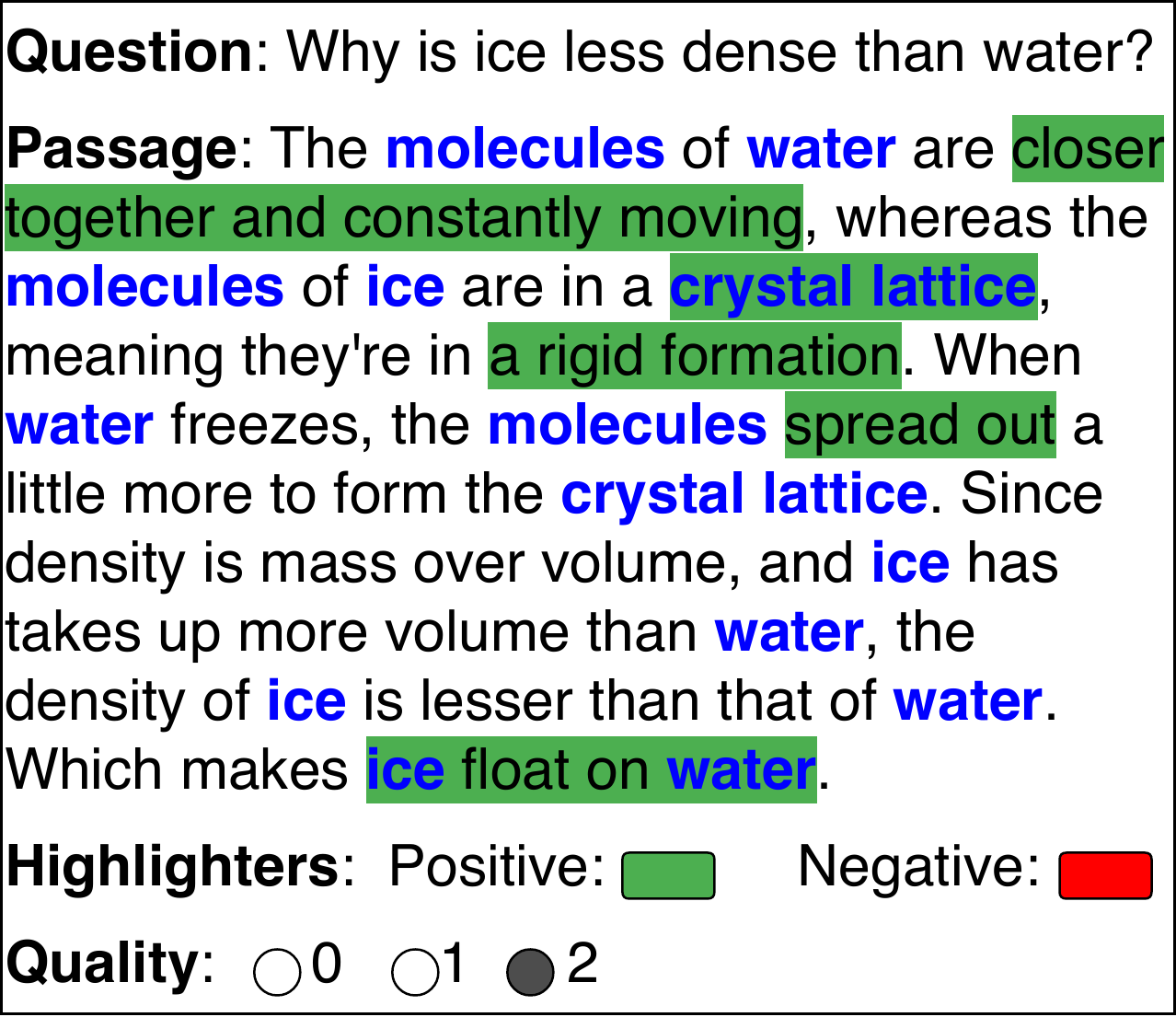}
  \vspace{-0.6cm}
	\caption{An illustration of the highlighting interface }
	\label{fig:interface-example}
\end{minipage}
\end{figure}

Turkers are instructed to rate the answer quality after highlighting the passage. Figure~\ref{fig:passage-highlight-ratings} presents the histogram of rated answer quality. 
We observe that turkers typically rate 1 or 2 for good answers and 0 for bad answers.
This indicates that \textit{turkers do not hesitate to rate a bad answer, but they can be severe on the answer quality judgments even in some cases where the passage is the answer}.

\begin{figure}[ht]
\centering
\begin{minipage}{0.233\textwidth}
  \centering
  \includegraphics[width=1\linewidth]{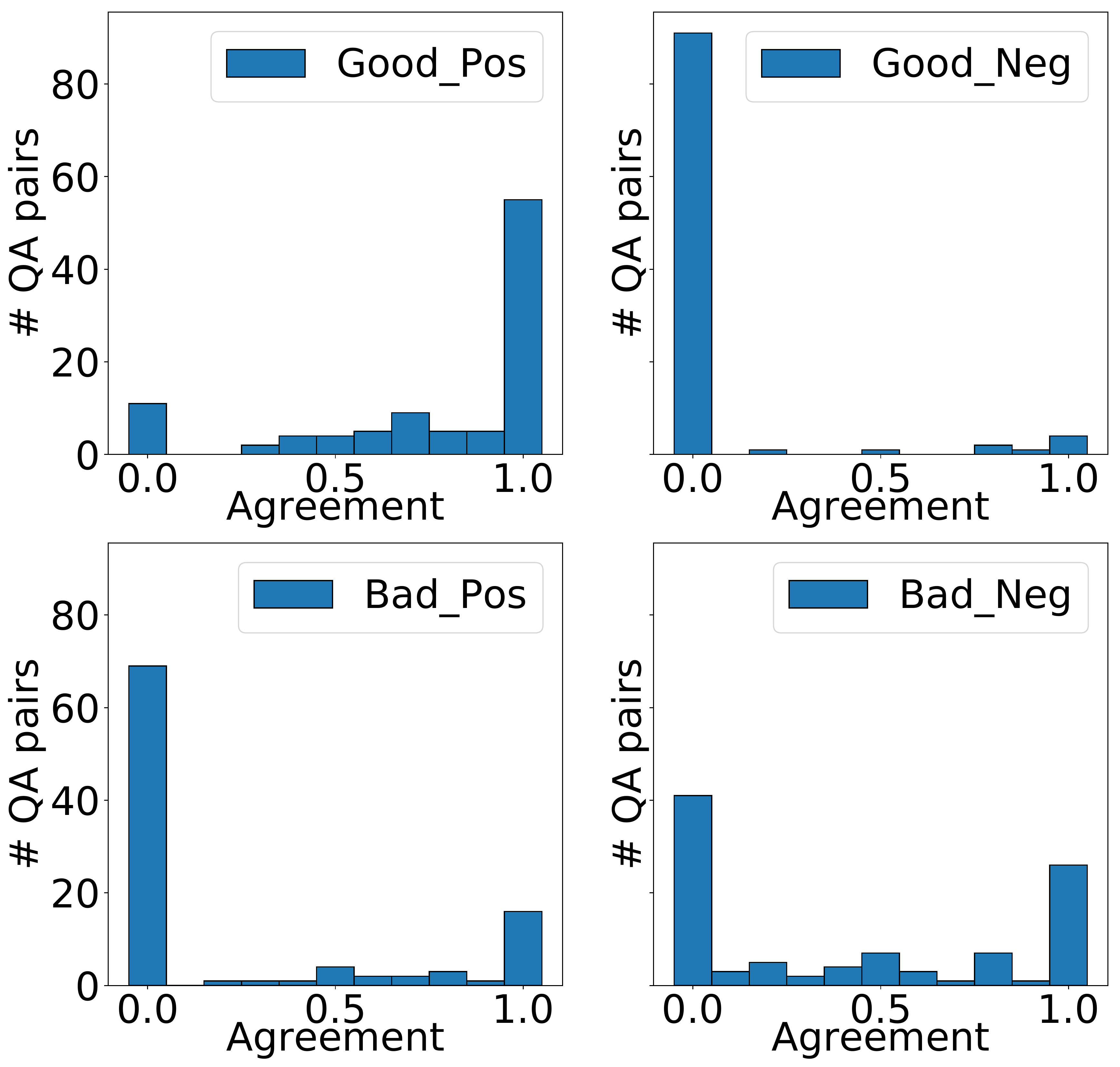}
  \vspace{-0.85cm}
    \caption{Agreement for the second setting}
	\label{fig:passage-highlight-agreement}
\end{minipage}%
\hspace{0.1cm}
\begin{minipage}{0.233\textwidth}
  \centering
  \includegraphics[width=1\linewidth]{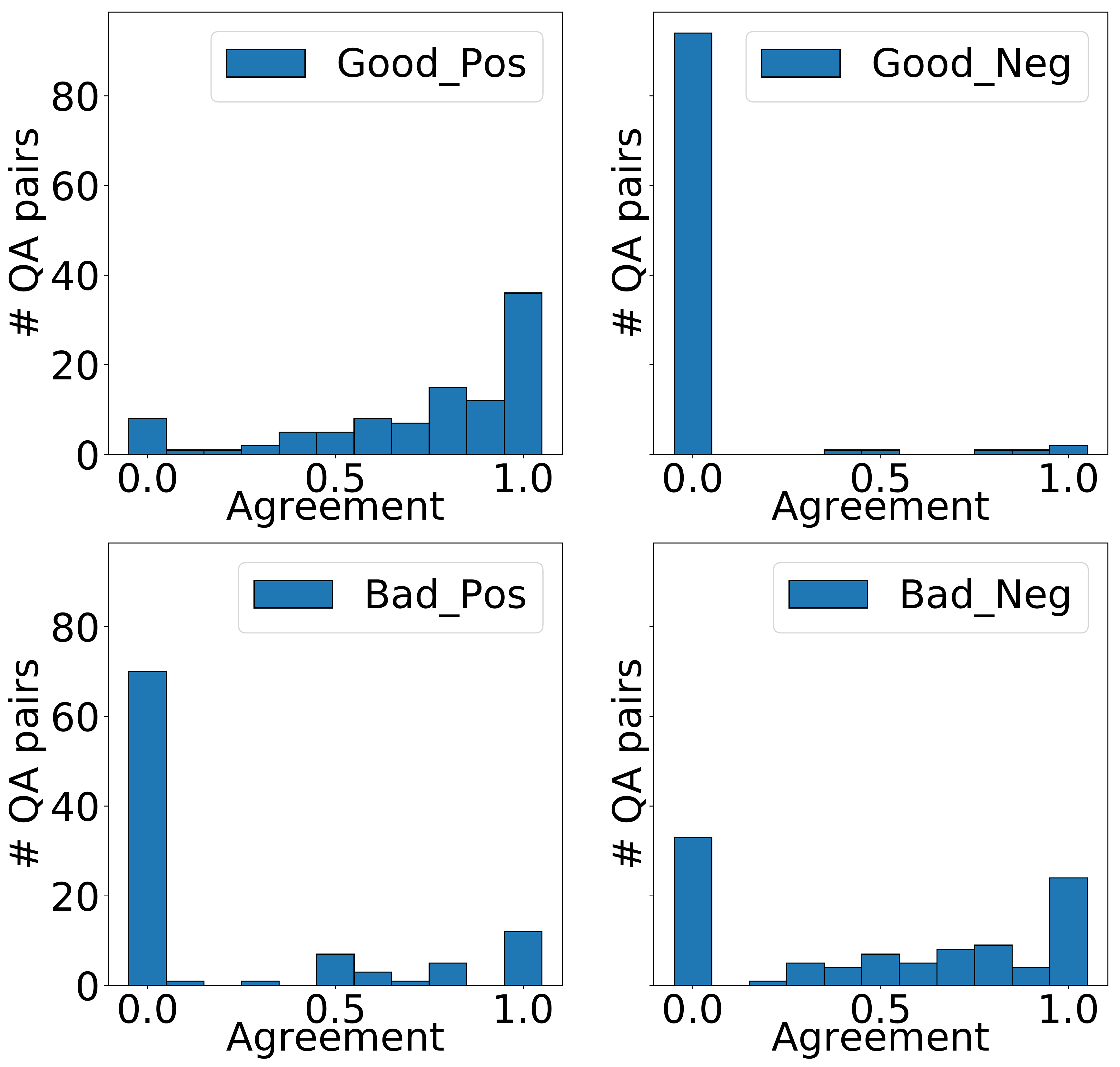}
  \vspace{-0.85cm}
	\caption{Agreement for the third setting}
	\label{fig:passage-highlight-2-agreement}
\end{minipage}
\end{figure}

On average, turkers highlight 7.57 words in good answers and 6.35 words in bad answers. In addition, the consistency rule adds about 0.5 words in both cases. Figure~\ref{fig:passage-highlight-agreement} presents the agreement of highlights in four different cases: positive highlight on good answers, positive highlight on bad answers, negative highlight on good answers, and negative highlight on bad answers. 

We notice the low agreement in Figures ``Good\_Neg'' and ``Bad\_Pos''. This can be accounted for by the fact that turkers rarely use negative highlights for good answers or positive highlights for bad answers. When one or both turkers in an agreement pair do not use a certain type of highlight (positive or negative), the agreement is 0 (turkers need to make at least one highlight, but they do not have to use both types of highlights). Therefore, we focus on positive highlights on good answers and negative highlights on bad answers. In the former case, more than half of the QA pairs achieve complete agreement, which means one turker's annotation is either exactly the same with or a subset of the other's. In the latter case, the agreement is fuzzy but it shows a total agreement in about 30\% of the QA pairs. The results indicate that \textit{people tend to get good agreement on what makes a good answer good. In contrast, when deciding what makes a bad answer bad, people tend to have more diverse opinions while still managing to achieve some agreement}.

\subsection{Passage Highlighting with Suggested Words}
\label{subsec:passage-highlight-2}

In this setting, the distributions of answer quality ratings are very similar to the last setting, where the suggested words are unmarked.

We mark an average of 9.48 suggested words in special styles. The average number of highlighted words by turkers is 6.98 for good answers and 6.06 for bad answers, slightly smaller than the last setting.
Overall, the agreement for the last setting (Figure~\ref{fig:passage-highlight-agreement}) is more polarized while the agreement for this setting (Figure~\ref{fig:passage-highlight-2-agreement}) has some distribution weight in the middle. These results indicate that some turkers may consider it unnecessary to highlight the marked words even when they believe the words are positive or negative.

One of the goals of this setting is to observe turkers' behavior under the impact of suggested words. This could be done by quantifying the agreement between highlights and suggested words with overlap coefficient. Figure~\ref{fig:passage-highlight-overlap} and \ref{fig:passage-highlight-2-overlap} shows the results for Passage Highlight and Passage Highlighting with Suggested Words settings.
They indicate that \textit{turkers tend to base their decision more on the suggested words when these words are marked in special styles. This suggests that marking important words in answers could influence peoples' decision making process in answer quality evaluation.} 

\subsection{Answer Quality vs. QA Text Similarity}
\label{subsec:quality-vs-similarity}
Some QA systems use the text similarity of questions and answers to perform answer retrieval \cite{Cohen2016NonFactoidQA,Yang2016NonFactoidQA}. However, we show that answer quality is not the same as QA text similarity. We take the majority vote of overall ratings for each QA pair as the final answer quality rating. The rating data comes from the third setting (data from the second setting also gives a very similar result). To compute the text similarity between the question and answer in a QA pair, we obtain the tf-idf representations and the aggregated word embedding representations (the sum of the word embeddings in a passage). Then we calculate the cosine similarity for both representations respectively and compute their harmonic mean as the QA similarity measure. We plot the histogram of QA text similarity under each quality level in Figure~\ref{fig:quality-vs-similarity}.

We observe from the figure that the QA text similarity is relatively low in general for non-factoid QA. We further make three observations: (1) Quality level 2 has some high similarity values (such as 0.7 and 0.8), which are rare in quality level 1 and 0. (2) Quality level 0 has more low similarity (such as 0) QA pairs than the other two. (3) All three quality levels have large numbers of medium level similarities.
These results indicate that \textit{QA text similarity does not necessarily capture answer quality. They can be positively correlated when it comes to very similar or very different QA pairs. However, text similarity cannot determine answer quality effectively if the QA pair has medium level similarity.} For example, the QA pair shown in Figure~\ref{fig:interface-example} only has a text similarity of 0.56, but received quality level 2 ratings from all three turkers. We plan to investigate this in more detail in future work.

\section{Conclusions} 
\label{sec:conclusion}
In this paper, we studied three different fine-grained answer presentation methods in a non-factoid QA setting. 
We also discovered that QA text similarity does not necessarily capture answer quality in this setting. Our findings are based on crowdsourcing and thus need to be generalized with caution. Our findings can be used in designing a more interactive IR system that emphasizes answer retrieval. Future work will include verifying our findings and exploring other answer interaction methods in a conversational setting.

\begin{acks}

This work was supported in part by the Center for Intelligent Information Retrieval, NSF IIS-1715095 and ARC DP180102687. Any opinions, findings and conclusions or recommendations expressed in this material are those of the authors and do not necessarily reflect those of the sponsor. 

\end{acks}

\bibliographystyle{ACM-Reference-Format}
\bibliography{acmart} 

\appendix
\section{Additional Figures}
\begin{figure}[ht]
\centering
\begin{minipage}{0.231\textwidth}
  \centering
  \includegraphics[width=1\linewidth]{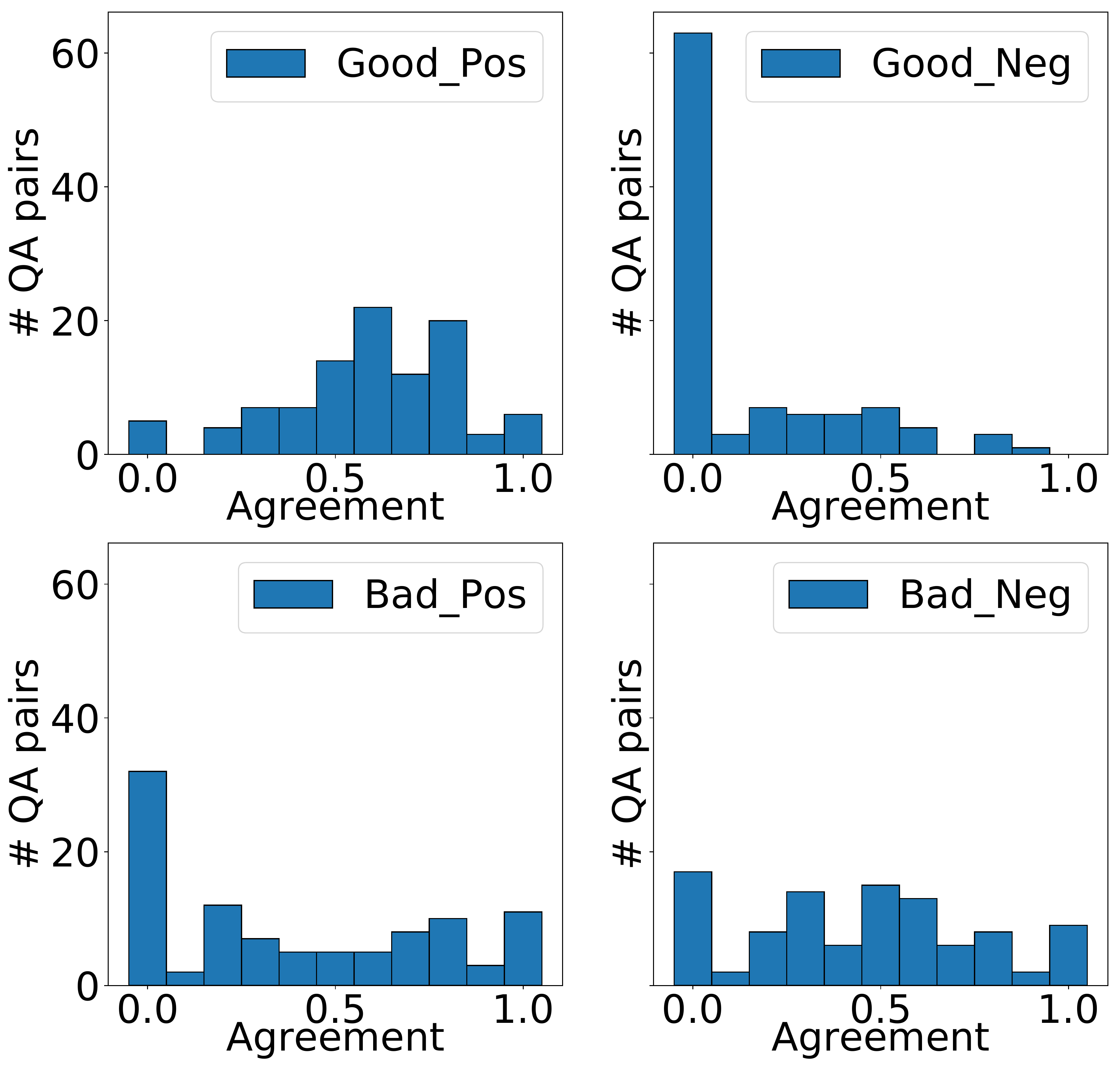}
    \caption{Suggestion overlap for the second setting}
	\label{fig:passage-highlight-overlap}
\end{minipage}
\hspace{0.1cm}
\begin{minipage}{0.231\textwidth}
  \centering
  \includegraphics[width=1\linewidth]{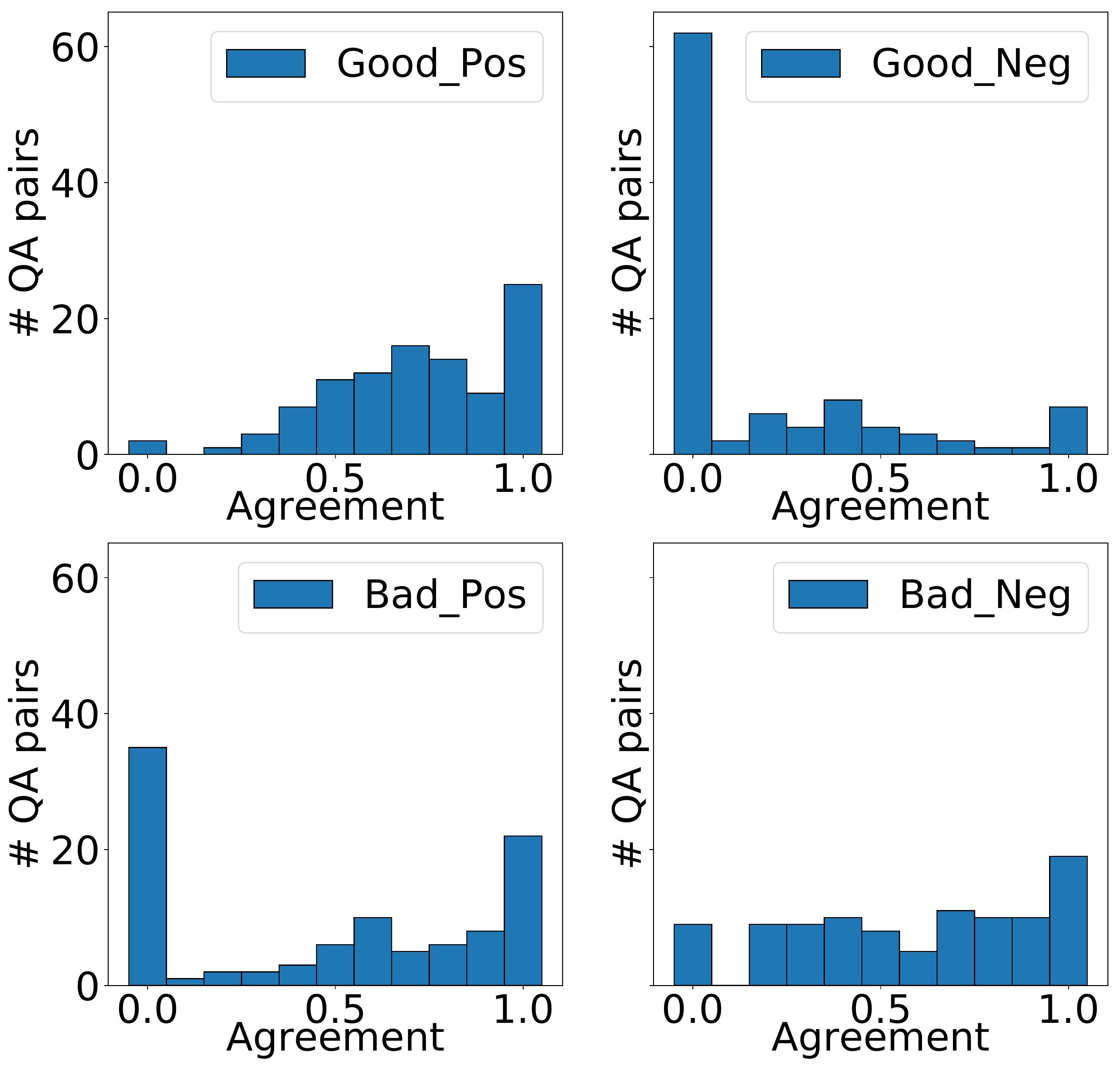}
	\caption{Suggestion overlap for the third setting}
	\label{fig:passage-highlight-2-overlap}
\end{minipage}
\end{figure}

\begin{figure}[ht]
    \centering
    \includegraphics[width=1\linewidth]{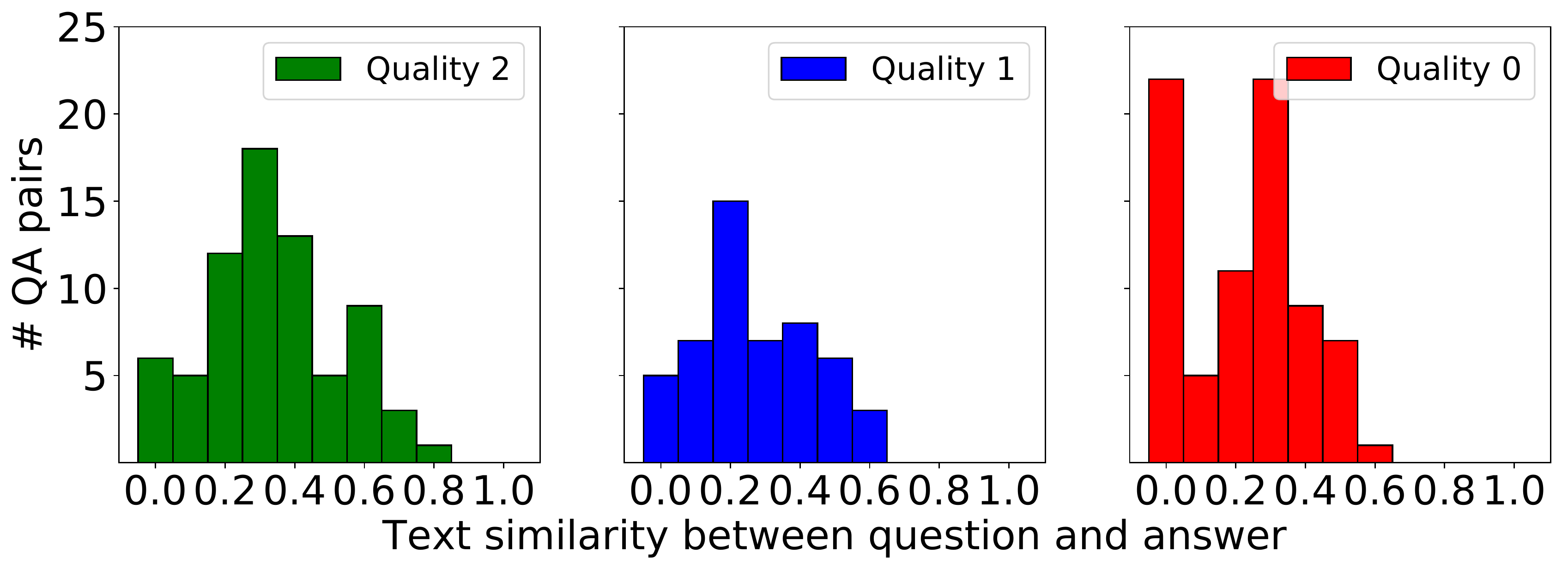}
    \caption{QA text similarity under each quality level}
    \label{fig:quality-vs-similarity}
\end{figure}

We include some additional figures to illustrate our findings from above. Figures~\ref{fig:passage-highlight-overlap} and \ref{fig:passage-highlight-2-overlap} shows the agreement between turkers' annotations and the suggested words. The two figures correspond to the Passage Highlight and Passage Highlighting with Suggested words settings respectively. In addition, we plot the histogram of QA text similarity under each quality level in Figure~\ref{fig:quality-vs-similarity}.

\end{document}